\documentclass[aps,prl,showpacs,twocolumn,superscriptaddress]{revtex4}

\usepackage{mathrsfs}
\usepackage{graphicx}
\usepackage{color}
\usepackage{amsmath}
\usepackage{amssymb}

\begin{document}

\title{Signal focusing through active transport}

\author{Alja\v{z} Godec}
\affiliation{Institute of Physics \& Astronomy, University of Potsdam, D-14476
Potsdam-Golm, Germany}
\affiliation{National Institute of Chemistry, 1000 Ljubljana, Slovenia}
\author{Ralf Metzler}
\affiliation{Institute of Physics \& Astronomy, University of Potsdam, D-14476
Potsdam-Golm, Germany}
\affiliation{Department of Physics, Tampere University of Technology, FI-33101
Tampere, Finland}

\begin{abstract}
In biological cells and novel diagnostic devices biochemical receptors need to
be sensitive to extremely small concentration changes of signaling molecules.
The accuracy of such molecular signaling is ultimately limited by the counting
noise imposed by the thermal diffusion of molecules. Many macromolecules and
organelles transiently bind to molecular motors and are then actively
transported. We here show that a random albeit directed delivery of signaling
molecules to within a typical diffusion distance to the receptor reduces the
correlation time of the counting noise, effecting an improved sensing precision.
The conditions for this \emph{active
focusing\/} are indeed compatible with observations in living cells. Our results
are relevant for a better understanding of molecular cellular signaling and
the design of novel diagnostic devices.
\end{abstract}


\maketitle 

Cellular signaling relaying external or internal biochemical cues typically
involves low copy numbers of messenger molecules, inevitably effecting
appreciable fluctuations in the count of molecular binding events at specific
receptors \cite{Bialek,BergPurc,Bialek2,Endres,Levine3,Tkacik,Levine4,Levine5,
Levine6,Govern,Wolde1,Tkacik2}. A similar limitation by counting noise is
encountered in modern microscopic diagnostic devices to which sensitivity is a
key factor \cite{Diagnostics}. Modern microscopic techniques reveal molecular
signaling events and underline their inherent stochasticity in living cells
\cite{Stoch,Xie,Elf1,Elf2}. Nevertheless molecular signaling pathways in
biological cells operate at remarkable precision \cite{BialekBK,Alberts}. 

The first heuristic argument about noise limitation to biological concentration
measurements is due to Berg and Purcell assuming biochemical receptors to count
the number $N$ of specific molecules in a volume equal to their linear dimension
$a$ \cite{BergPurc}. $N$ is then limited by Poissonian noise $\delta N\sim\langle
N\rangle^{1/2}$. The time between two independent measurements is set by the time
$\tau_D\sim a^2/D$ needed to clear the volume by diffusion, $D$ being the molecular
diffusivity. Averaging over a time $\tau_m$ thus allows to $N_m\sim\tau_m/\tau_D$
independent measurements, reducing the noise by the factor $N_m^{1/2}$. The
relative accuracy to measure a background concentration $\langle c\rangle$ is thus
$\overline{\delta c^2}/\langle c\rangle^2\sim(Da\langle c\rangle\tau_m)^{-1}$
\cite{BergPurc}. When the additional binding dynamics to the biochemical receptor
is explicitly taken into account, this relative error becomes \cite{Bialek}
\begin{equation}
\label{BandS}
\frac{\overline{\delta c^2}}{\langle c\rangle^2}=\frac{2}{k_{\mathrm{on}}\langle
c\rangle(1-\langle n\rangle)\tau_m}+\frac{1}{\pi Da\langle c\rangle\tau_m}.
\end{equation}
The first contribution stems from the Markovian (un)binding to the receptor at
detailed balance conditions with
binding rate $k_{\mathrm{on}}$ and average receptor occupancy $\langle n\rangle$.
The second term is the diffusional
noise, up to the factor $\pi$ identical to the result by Berg and Purcell
\cite{BergPurc}. The prefactor of the diffusive term in Eq.~(\ref{BandS}) was
recently refined heuristically \cite{Wolde1}. Inspired by the early ideas of
Berg and Purcell a number of studies unraveled the crucial role of diffusional
noise in biochemical signaling \cite{Endres,Levine3,Levine4,Levine5,Levine6,Govern}
along with additional features such as receptor cooperativity \cite{Bialek2}
and facilitated diffusion \cite{Tkacik2}. Various experiments suggest that
cells indeed operate very close to the fundamental accuracy limit \cite{Bialek}.  

Here we extend the result (\ref{BandS}) to the case when the signaling molecules
are not only freely diffusing in the cell but actively transported along cellular
filaments by motor proteins \cite{Motors1,Motors2} effecting intermittent ballistic
excursions \cite{Interm1,Interm2,Interm3,Heinrich}. Such an additional active
component occurs when extracellular signaling molecules are taken up by the cell
via endocytosis: the molecules are engulfed into submicron lipid vesicle and then
intermittently transported through the cell by motors \cite{endocytosis}.
A similar combination of free diffusion and active transport occurs when virus
particles invade a living cell \cite{seisenhuber}. However, even free molecules
such as messenger RNA may attach to motors \cite{mrna}, or proteins move in
directed fashion due to cytoplasmic drag \cite{haim}. To incorporate the active
component we employ the theory of random intermittent search for hidden targets
\cite{RMP} which was recently used to analyze reaction kinetics in active
media \cite{Olivier}. We show that active transport enables both faster
as well as more accurate sensing: an active noise floor exists, but it can be
significantly lower than the purely diffusive counting noise (\ref{BandS}). This
\emph{active focusing\/} reduces the noise correlation time and enables the
receptor to detect relative changes in concentration with higher accuracy. Our
results also have direct implications to the design of active components in
microscopic synthetic diagnostic systems based on molecular signals
\cite{Diagnostics}.

\begin{figure}
\includegraphics[width=8.cm]{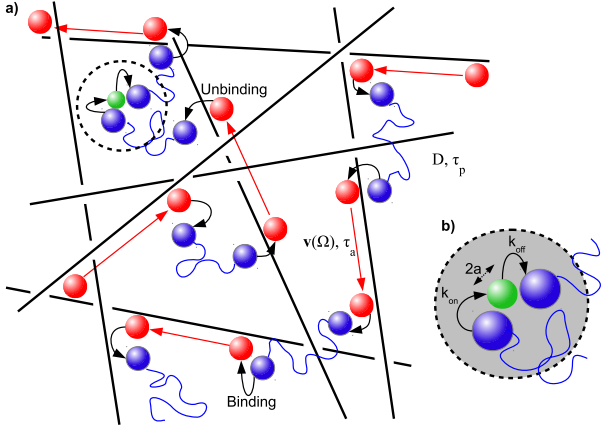}
\caption{Schematic of the model system: a) Each signaling particle performs
passive thermal diffusion (blue phases) interrupted by active ballistic
excursions with constant speed and random direction (red phases moving
along the black motor tracks). The duration of both phases is distributed
exponentially with mean times $\tau_{p,a}$. When the particle reaches the
receptor (green sphere) it binds with on-rate $k_\mathrm{on}$ and dissociates
with rate $k_\mathrm{off}$. b) Magnification of the receptor region.}
\label{schm}
\end{figure}

\emph{Model.} We consider a signaling particle (vesicle, virus, mRNA, or protein)
in 3-dimensional cellular space, randomly switching between a passive diffusion
phase $p$
with diffusivity $D$ and an active ballistic phase $a$ with velocity $\mathbf{v}
(\Omega)$ of constant magnitude $v=|\mathbf{v}|$ \cite{REM} in the direction of the
solid angle $\Omega$ following a Markovian dynamics (Fig.~\ref{schm}). Assuming
ideally disordered cytoskeletal filament orientations, the spatial direction of
active motion events is uniformly distributed. The duration
of active/passive phases is exponentially distributed with mean $\tau_{a,p}$
\cite{Arg}. The concentrations of freely diffusing and motor-bound signal particles
are $c_p(\mathbf{r},t)$ and $c_a(\mathbf{r},\Omega,t)$. A receptor with radius $a$
is placed at $\mathbf{r}_0$. Then the fractional occupancy $n(t)$ of the receptor
by a signal particle evolves according to a mean field kinetic scheme obeying
detailed balance with on/off rates $k_\mathrm{on/off}$,
\begin{equation}
dn(t)/dt=k_{\mathrm{on}}c_{p}(\mathbf{r}_0,t)(1-n(t))-k_{\mathrm{off}}n(t).
\end{equation}
Assuming that the particle (un)binds to (from) the receptor only from (to) the 
passive mode \cite{Olivier}, the coupled set of equations for the
concentrations $c_{a,p}$ reads
\begin{eqnarray}
\nonumber
\label{model}
&&\frac{\partial c_{p}(\mathbf{r},t)}{\partial t}=D\nabla^2c_p+\int\frac{c_a}{
\tau_a}d\Omega-\frac{c_p}{\tau_p}-\delta(\mathbf{r}-\mathbf{r}_0)\frac{dn(t)}{dt},\\
&&\frac{\partial c_a(\mathbf{r},\Omega,t)}{\partial t}=-\nabla_{\mathbf{r}}\cdot
(\mathbf{v}(\Omega)c_a)-\frac{c_a}{\tau_a}+\frac{c_p}{4\pi\tau_p}.
\end{eqnarray}

The signaling typically occurs in two stages. In the initial phase a change in
the concentration of the signaling particles occurs either by exchange with the
extracellular space \cite{Alberts,Bressloff} or by variation of the production
and/or degradation rates. Upon re-equilibration (assumed to be much faster than
the measurement time $\tau_m$) the receptor reads out the concentration over the
time $\tau_m$ in the \emph{measurement phase}. In a diagnostic device equivalent
phases will be observed after sample immersion and during detection periods.
In an optimal signaling setup equilibration should be as fast as possible while
the measurement phase should be as precise as possible. We now quantify the
\emph{speed} and \emph{precision} of the two signaling phases.

\begin{figure*}
\includegraphics[width=16.cm]{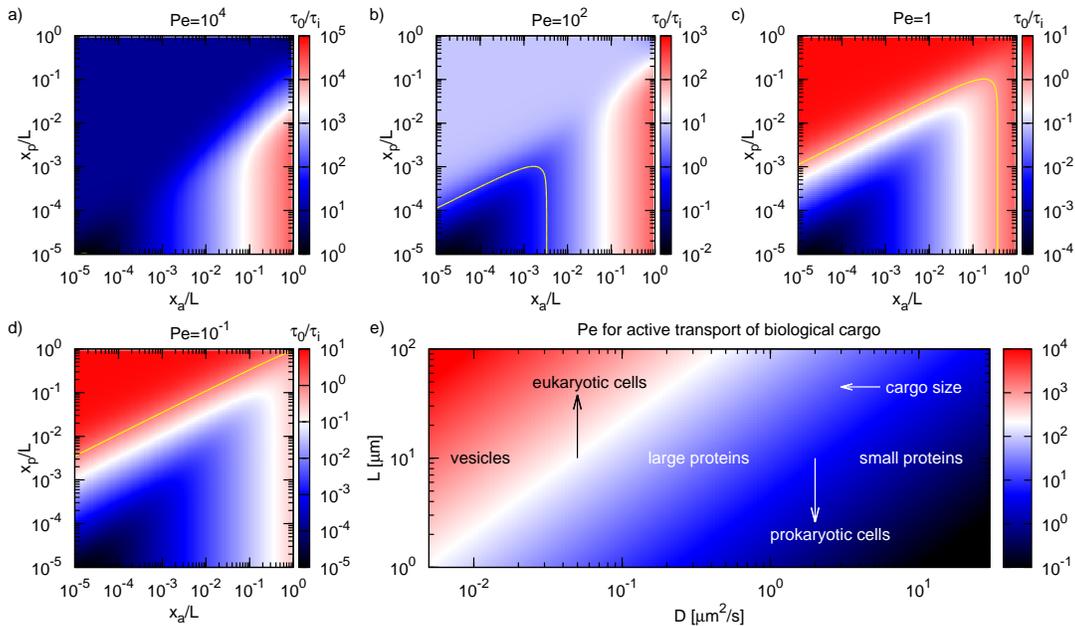}
\caption{a)-d) Equilibration times $\tau_0/\tau_i$ for various P{\'e}clet numbers
$Pe$ from numerical inversion of Eq.~(\ref{msd}). Yellow lines denote $\tau_0/\tau
_i=1$. When $\tau_0/\tau_i>1$ active trafficking is more efficient. Note the
different scales for $\tau_0/\tau_i$. e) $Pe$ as
function of cell (domain) size $L$ and diffusivity $D$ for biological cells
\cite{Alberts,Bressloff}. Arrows denote increasing particle size from right
(small proteins) to left (large vesicles) and cell size from bottom (prokaryotic
cells) to top (eukaryotic cells). The
border between prokaryotic and eukaryotic cells is around
10$\mu$m. For $Pe\sim10^4$ (vesicles or large proteins in
eukaryotic cells), active transport always improves the speed. For
small proteins in eukaryotic cells and large
proteins in prokaryotic cells ($10\lesssim Pe\lesssim10^2$), in addition to large
$x_a$ a large $x_p$ is necessary for active transport to be more efficient. For
$Pe\lesssim10^{-1}$ active transport is always less efficient.}
\label{Speed}
\end{figure*}

\emph{Speed.} We assume that the system equilibrates on a time scale over which
the signaling molecules move a distance $L$ of the order of the cell size (or
that of a cellular compartment). At this stage we neglect the analyte-receptor
binding dynamics and adopt a probabilistic view of Eqs.~(\ref{model}). The
equilibration time
$\tau_i$ is then defined by the mean squared displacement (MSD), $\langle|\mathbf{
r}(\tau_i)|^2\rangle=L^2$. The exact result is (see Supplementary Material, SM)
\begin{eqnarray}
\nonumber
&&\hspace*{-0.6cm}
\langle |\mathbf{r}^2(t)|\rangle=2\Bigg\{(v\tau_a)^2e^{-\frac{t}{\tau_a}}-
\frac{v^2+3D\tau_p^{-1}}{(\tau_a^{-1}+\tau_p^{-1})^2}e^{-\frac{t(\tau_a+\tau_p)}{
\tau_a\tau_p}}\\
&&\hspace*{-0.6cm}
+\frac{\tau_p^{-1}(v\tau_a)^2+3D}{1+\tau_a/\tau_p}t+\frac{3D\tau_p-(v\tau_a)^2
(1+2\tau_p/\tau_a)}{(1+\tau_p/\tau_a)^2}\Bigg\}.
\label{msd}
\end{eqnarray}
Over a period of duration $\tau_a+\tau_p$, during which the directional persistence
in the active phase causes a nonlinear time dependence of $\langle|\mathbf{r}^2(t)|
\rangle$ and hence a local violation of the central limit theorem, an effective
diffusive regime $\langle|\mathbf{r}^2(t)|\rangle\simeq t$ is established.
Eq.~(\ref{msd}) is a transcendental equation for $\tau_i$, essentially depending
on only three parameters: the typical distance covered in the active and passive
phases, $x_a=v\tau_a$ and $x_p=\sqrt{D\tau_p}$, and the P{\'e}clet number $Pe=Lv/D$.
To estimate the efficiency of active trafficking with respect to diffusion we
compare $\tau_i$ with the purely passive equilibration time $\tau_0\equiv L^2/(6D)$.
The results for various $Pe$ values (see Fig.~\ref{Speed}e)) typical for
biological systems are shown in Fig.~\ref{Speed}a)-d).

Active transport is more efficient for larger particles (small $D$) in larger
domains,
a direct consequence of the finite motor velocity and instantaneous directionality
of active motion. Namely, in terms of the MSD diffusion and active motion display
different time scaling ($\simeq t$ versus $\simeq t^2$): considering only pure
passive and active motion for $Pe<6$ diffusion is more efficient. In the
intermittent case the motion has a transient period of duration $\tau_a+\tau_a$,
which corresponds to a parameter-dependent combination of both regimes during the
relaxation towards the equilibrium partitioning between phases $a$ and $p$. After
this transient period an effective diffusive regime is established with diffusivity
$D_{\mathrm{eff}}=(D+x_a^2/(3\tau_p))/(1+\tau_a/\tau_p)$, which may or may not be
larger than $D$. $\tau_i$ can thus be smaller or larger than $\tau_0$. Trafficking
of vesicles with $D\lesssim10^{-2}\mu\mathrm{m}^2/\mathrm{s}$ therefore mostly
profits from active motion, whereas active motion of smaller proteins with $D\sim
10\mu\mathrm{m}^2/\mathrm{s}$ will only be more efficient over large distances as
in eukaryotic cells (especially for neurons), and only if accompanied by
significant phases of passive diffusion. The observed features explain why it is
profitable for a cell to use active transport for trafficking of larger particles
\cite{BialekBK,Kinesins}, despite demanding more cellular resources.
Similarly, active diagnostics \cite{Diagnostics} can be faster and hence allow for
a higher throughput.

\emph{Precision.} Since the precision of the receptor measurement of the signal
molecule concentration should be maximized we consider small deviations from the
equilibrium values $n=\langle n\rangle+\delta n$ and $c_{p,a}=\langle c_{p,a}\rangle
+\delta c_{p,a}$ \cite{Bialek},
taking into account the detailed balance for the binding/unbinding transitions via
$k_{\mathrm{on}}\langle c_p\rangle/k_{\mathrm{off}}=\exp(F/[k_BT])$, where $F$ is
the binding free energy \cite{supp}. Solving Eqs.~(\ref{model}) and following the
ideas of Ref.~\cite{Bialek} we use the fluctuation-dissipation theorem to obtain
the power spectrum of the concentration fluctuations. Within the linear
response regime and for receptor measurement times $\tau_m$ exceeding any
correlation times, the relative error for the concentration measurement is
\cite{supp}
\begin{equation}
\frac{\overline{\delta c_p^2}}{\langle c_p \rangle^2}=\frac{2}{k_{\mathrm{on}}
\langle c_p\rangle(1-\langle n\rangle)\tau_m}+\frac{\Lambda(x_a,x_p)}{\pi Da
\langle c_p\rangle\tau_m}.
\label{sigma}
\end{equation}
Compared to Ref.~\cite{Bialek} for passive diffusion our result (\ref{sigma})
differs by the dimensionless factor $\Lambda(x_a,x_p)$ in the second term, for
which always $\Lambda(x_a,x_p)\le1$ holds \cite{supp}.

We now focus on the transport controlled regime in which $k_{\mathrm{on}}
\langle c_p\rangle,k_{\mathrm{off}}\gg\tau_a^{-1},\tau_p^{-1}$ and $k_{\mathrm{on}}
\langle c_p\rangle/k_{\mathrm{off}}
\gg 1$. Hence we neglect the first term in Eq.~(\ref{sigma}) and consider the
remaining \emph{active noise floor}. If the active excursions are short compared
to the receptor size, $x_a/a\ll1$,
\begin{equation}
\label{asympt1}
\Lambda(x_a,x_p)\sim\frac{\sqrt{5}ax_p}{(\pi x_a)^2}\frac{\tanh^{-1}\left(\frac{
(\pi x_a)^2}{\sqrt{5}ax_p}\left[1+\frac{1}{3}(x_a/x_p)^2\right]^{-\frac{1}{2}}
\right)}{[1+\frac{1}{3}(x_a/x_p)^2]^{1/2}}. 
\end{equation}
In the biologically more important situation $x_a\gg a$,
\begin{equation}
\label{asympt2}
\Lambda(x_a,x_p)\sim1-\tan^{-1}(\pi x_p/a)/(\pi x_p/a),
\end{equation}
which has the lower bound $\Lambda_{\mathrm{min}}\sim(\pi x_p/a)^2[1-(\pi x_p/a)^2
]$ as $x_p/a\to0$. This might suggest an approach towards an infinite absolute
precision of the transport term as $x_p\to0$. However, at fixed total concentration
$c_{\mathrm{tot}}$ of signaling particles in the transport controlled regime we
have that $\langle c_p\rangle=c_{\mathrm{tot}}/(1+\tau_a/\tau_p)$ corresponding to
$\langle c_p\rangle\to0$ as $x_p\to0$, hence diverging relative fluctuations.
Therefore, there still exists a noise floor to active sensing but it can be
significantly reduced as explained below.

\begin{figure*}
\includegraphics[width=16.cm]{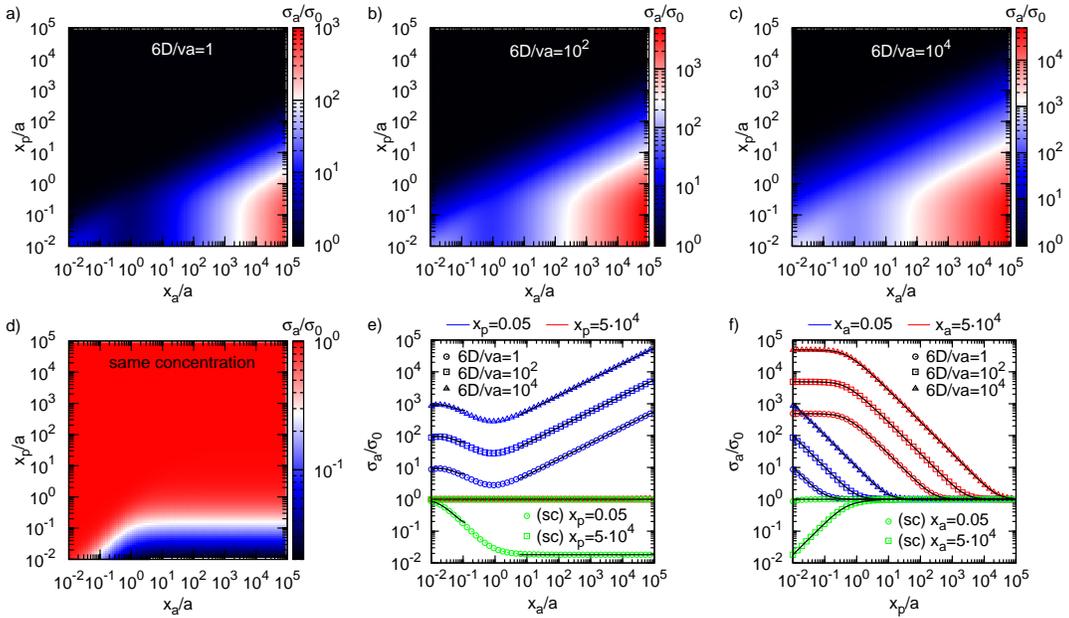}
\caption{a)-c) Fractional variance $\sigma_a=[\overline{\delta c_p^2}/\langle c_p
\rangle^2]^{1/2}$ of concentration fluctuations at the receptor site for active
trafficking compared to thermal diffusion, $\sigma_0=[\pi Dac_{\mathrm{tot}}\tau_m]
^{-1/2}$ as function of the average distance traveled in both phases in units of the
receptor radius $a$. In a)-c) the total concentration $c_{\mathrm{tot}}$ is kept
constant, hence decreased sensing accuracy is solely due to exceedingly small
equilibrium concentrations in the passive phase. d) $\sigma_a/\sigma_0$ for equal
equilibrium concentrations (sc in parts e), f)) in the passive phase $\langle
c_p\rangle$ for different total concentrations $=c_{
\mathrm{tot}}(1+\tau_a/\tau_p)$: a strong increase in sensing accuracy is
observed for $x_p/a\to0$. a)-d) use the full results (\ref{sigma}) and
(\ref{exact}) in SM \cite{supp}. e) Horizontal cross-sections of $\sigma_a/\sigma_0$
at large and
small $x_p$ (symbols) compared to the approximations (\ref{asympt1}),
(\ref{asympt2}). f) Vertical cross-sections at large and small $x_a$ (symbols)
compared to the approximations (\ref{asympt1}), (\ref{asympt2}). Note
the different scales for $\sigma_a/\sigma_0$ and the excellent agreement between
full and approximate results. The non-monotonicity at small $x_p$ is due to
the interplay of signal focusing (smaller $\Lambda$) and
decreasing $\langle c_p\rangle$.}
\label{vars}
\end{figure*}

We gauge the sensing precision at equal $c_{\mathrm{tot}}$ and equal $\langle c_p
\rangle$, corresponding to the lower and upper bounds for the gain of active
focusing. For equal $c_{\mathrm{tot}}$ we compare the relative accuracy of
measuring \emph{different\/} concentrations of freely diffusing molecules
in active and passive sensing, finding that the precision is always worse for
active transport as compared to free diffusion (Fig.~\ref{vars}a)-c) and
asymptotic results in e) and f)) and becomes worse with longer active excursions.
The reason is that despite reducing the absolute fluctuations by active dynamics
we are measuring smaller and smaller effective concentrations.
Conversely, Figs.~\ref{vars}d)-f) shows that
if we compare the precision at identical concentration
$c_{\mathrm{tot}}(1+\tau_a/\tau_p)$ of free molecules the accuracy can be improved
significantly as long as $x_p\lesssim a$ (otherwise, in this regime active
trafficking does not affect sensing precision). The ingredients necessary to
understand this reduced counting noise are: (i) only fluctuations on a scale
$\sim a$ are relevant for the sensing accuracy and (ii) the receptor only 'sees'
free particles. Hence, at finite temperatures perfect signaling corresponds to
the situation in which upon release from the receptor the particles immediately
bind to a motor and are swept away over distances $>a$. Concurrently diffusive
displacements must be $\lesssim a$ to assure focused delivery in the sense that
any unbinding form the motor only contributes if it occurs at $|\mathbf{r}-
\mathbf{r}_0|\lesssim a$. This reduction of local concentration fluctuations by
means of intermittent active excursions is exactly the {\em active signal
focusing\/} mentioned above.

\emph{Conclusion.} The last years have seen significant activity to explore the
counting noise for purely diffusive scenarios \cite{Bialek,BergPurc,Bialek2,
Endres,Levine3,Tkacik,Levine4,Levine5,Levine6,Govern,Wolde1} and to analyze the
speedup of receptor binding due to the topological coupling of one- and
three-dimensional diffusion for gene regulation in the facilitated diffusion model
\cite{Otto,Tkacik2}. In this Letter we fill the apparent gap in the quantitative
assessment of the counting noise experienced by biochemical receptors measuring
the local concentration of signaling molecules or compounds in the case when an
active transport component is present. This occurs for various signaling
particles (proteins, mRNA molecules, vesicles containing signaling molecules, or
viruses) by direct shuttling of these signaling cues by molecular motors or by
cytoplasmic drag.

Compared to the purely diffusive signaling considered so far we showed that the
counting noise (the limit to accurate receptor measurement) for active sensing can
become significantly reduced due to active focusing. The only contributions to the
counting noise stem from particles, which are actively transported to within the
particle's typical free diffusion distance to the receptor. This reduces the
correlation time of the receptor occupancy noise and renders the averaging over a
measurement time $\tau_m$ more efficient. The importance of active signaling in
cellular regulation is well recognized \cite{ActSignal}. In agreement with our
results, in biological systems active transport is indeed employed to move larger
particles (e.g., vesicles or viruses) with intrinsically small $D$ \cite{Alberts,
Interm1,Interm2,Interm3,Kinesins,seisenhuber}. As a result even during longer
periods of detachment from motors these particles barely move \cite{BialekBK,
Interm1,Interm2,Interm3,Kinesins}. The typical experimental values $x_a\simeq
0.5-10\mu$m  \cite{Interm2,Interm3, Heinrich,Kinesins} for $a\simeq 1-10$nm in
fact fulfill the requirements of our model for signal focusing. However, as
discussed here also smaller particles such as mRNA and proteins experience active
motion components \cite{mrna,haim}, effecting active focusing for their detection.
In living cells the motor tracks are often not ideally disordered, as assumed
here, but biased towards the receptor \cite{ActSignal}. An expected net directional
bias towards the receptor, while not impeding signal focusing as long as $x_a\gg a$
and $x_p\lesssim a$ \cite{Arg}, would enhance the rate of delivery and
simultaneously increase the local concentration $\langle c_p(\mathbf{r}_0)\rangle$
at the receptor for equal $c_{\mathrm{tot}}$. Signal focusing is thus inherent to
active cellular signaling. Conversely, despite the great technological advance
over the past years, molecular motor-powered diagnostic devices have not yet
demonstrated a performance beyond the existing passive techniques, but a large
superiority is much anticipated \cite{Diagnostics}. Our results confirm
these expectations and present a first rigorous theoretical basis for their
systematic improvement and development.

\acknowledgments

We thank Andrey Cherstvy, Olivier B{\'e}nichou, and Raphael
Voituriez for discussions. We acknowledge funding through an Alexander von Humboldt
Fellowship (to AG) and an Academy of Finland FiDiPro grant (to RM).

\clearpage

\onecolumngrid

\renewcommand{\theequation}{S\arabic{equation}}
\setcounter{equation}{0}

\begin{center}
\textbf{Supplementary material\\ Signal focusing through active transport}\\[0.2cm]
Alja\v{z} Godec$^{\small 1,2}$ and Ralf Metzler$^{\small 1,3}$\\
\emph{$^1$Institute of Physics \& Astronomy, University of Potsdam, D-14476
Potsdam-Golm, Germany\\
$^2$National Institute of Chemistry, 1000 Ljubljana, Slovenia\\
$^3$Department of Physics, Tampere University of Technology, FI-33101
Tampere, Finland}\\[0.6cm]
\end{center}

\noindent
In this Supplemental Material we summarise the calculations and full results for
active transport coupled to reversible binding to a receptor.

\section{Speed} 

In this part we neglect the binding to the receptor (see
main text for argumentation) and invoke a probabilistic interpretation
of the model (i.e., concentrations $\to$ probability densities). By
Laplace transforming Eqs.~(2) and (3) in the main text in time and
Fourier transforming in space, $(\mathbf{r},t)\to(\mathbf{k},s)$ can be solved
for $\tilde{c}_a$ and
$\tilde{c}_p$. The Laplace image of the mean squared displacement
(MSD) for a particle starting at the origin in the passive phase is obtained from $\langle
|\mathbf{r}^2(s)|\rangle=-\nabla^2_{\mathbf{k}}(\int{\tilde{c}_ad\Omega}+\tilde{c}_p)|_{\mathbf{k}=0}$
and interchanging the order of differentiation with respect to
$\mathbf{k}$ and integration with respect to $\Omega$:
\begin{equation}
\langle
|\mathbf{r}^2(s)|\rangle=\frac{6D\left(1+\frac{\tau_p^{-1}}{s+\tau_a^{-1}}\right)+2\frac{v^2}{\tau_p}\frac{s+\tau_a^{-1}+\tau_p^{-1}}{(s+\tau_a^{-1})^3}}{s+\tau_p^{-1}-(\tau_a\tau_p)^{-1}/(s+\tau_a^{-1})},
\label{lmsd}
\end{equation}
which can be inverted exactly to give Eq.~(4) in the main text. The equilibration
time $\tau_i$ is defined implicitly by $\langle |\mathbf{r}^2(\tau_i)|\rangle=L^2$,
$L$ being the linear dimension of
the cell or its compartment. The transcendental equation for  $\tau_i$ is
solved numerically.

\section{Sensing precision} 
The Fourier transformed linearized Eqs. (1) read
\begin{eqnarray*}
\label{linmodel}
-i\omega
\delta\tilde{n}(\omega)&=&-\tau_c^{-1}\delta\tilde{n}(\omega)+k_{\mathrm{on}}(1-\langle
n\rangle)\delta\tilde{c_p}(\mathbf{r}_0,\omega)
 +k_{\mathrm{off}}\langle
n\rangle\beta\delta\tilde{F}(\omega)
\nonumber \\
-i\omega
\delta\tilde{c_a}(\mathbf{k},\Omega,\omega)&=&-(i\mathbf{v}(\Omega)\cdot\mathbf{k}+\tau_a^{-1})\delta\tilde{c_a}(\mathbf{k},\Omega,\omega)
+(4\pi \tau_p)^{-1}\delta\tilde{c_p}(\mathbf{k},\omega)
\nonumber\\
-i\omega
\delta\tilde{c_p}(\mathbf{k},\omega)&=&-(Dk^2+\tau_p^{-1})\delta\tilde{c_p}(\mathbf{k},\omega)+ \tau_{a}^{-1}\int\delta\tilde{c_a}(\mathbf{k},\Omega,\omega)d\Omega
+i\omega\delta\tilde{n}(\omega)e^{-i\mathbf{k}\cdot\mathbf{r}_0} 
\end{eqnarray*}
where we have used the constraint imposed by
detailed balance $\delta k_{\mathrm{on}}/k_{\mathrm{on}}-\delta
  k_{\mathrm{off}}/k_{\mathrm{off}}=\delta F/(k_BT)$ when linearizing
  the first of Eqs. (1) in the main text. Solving for
  $\delta\tilde{n}(\omega)$ we obtain 
\begin{equation}
\label{sol}
\delta\tilde{n}(\omega)=\frac{k_{\mathrm{off}}\langle
  n\rangle\beta\delta\tilde{F}(\omega)}{\tau_c^{-1}-i\omega \left(1+k_{\mathrm{on}}[1-\langle
    n\rangle](2\pi)^{-3}\int{d\mathbf{k}\Xi(\mathbf{k},\omega)}\right)},
\end{equation}
where $\beta=1/(k_BT)$ and we have defined the correlation time
$\tau_c\equiv(k_{\mathrm{on}}\langle c_p\rangle-k_{\mathrm{off}})^{-1}$ for
two-state Markovian switching \cite{BialekSM}
as well as the two auxiliary functions defined as
\begin{equation}
\label{aux} 
\Xi(\mathbf{k},\omega)\equiv-i\omega+Dk^2+\tau_p^{-1}-(4\pi\tau_p\tau_a)^{-1}\Psi(\mathbf{k},\omega)
\end{equation}
\begin{equation}
\Psi(\mathbf{k},\omega)\equiv\int_0^{\pi}d\theta\sin(\theta)\int_0^{2\pi}
\frac{d\varphi}{\tau_{p}^{-1}-i(\omega-\mathbf{v}(\theta,\varphi)\cdot\mathbf{k})}.
\end{equation}
We are interested in the low-frequency limit and hence take
$\Psi(\mathbf{k},\omega)\simeq \Psi(\mathbf{k},0)$. In this limit
the integral over $\varphi$ in $\Psi(\mathbf{k},\omega)$ can be evaluated as a contour integral along
the unit circle via the method of residues, while the second one is
amenable directly leading to 
\begin{equation}  
\Psi(\mathbf{k},0)=\frac{4\pi}{v k}\tan^{-1}\left(v\tau_ak\right),
\label{psiaux}
\end{equation}
where $k\equiv|\mathbf{k}|$. Using Eq. (\ref{psiaux}) we can obtain $\delta\tilde{n}(\omega)$
as a function of $\delta\tilde{F}(\omega)$ explicitly. The linear response function of the receptor occupancy (the
coordinate) to a change in the free energy difference between bound and freely diffusing
species (the thermodynamically conjugate force) is then
$\delta\tilde{n}(\omega)/\delta\tilde{F}(\omega)$ and is related to
the power spectrum of $n$ via the fluctuation-dissipation theorem
$S_n(\omega)=2/(\beta\omega)\mathrm{Im}[\delta\tilde{n}(\omega)/\delta\tilde{F}(\omega)]$. A
change in concentration is equivalent to a change in $F$ as a
consequence to a change in chemical potential, $\delta c_p/\langle
c_p\rangle=\beta\delta F$. Using this it can be shown \cite{BialekSM}
that $S_c(\omega)=(\beta  \langle
c_p\rangle)^2|\delta\tilde{n}(\omega)/\delta\tilde{F}(\omega)|^{-2}S_n(\omega)$
leading to
\begin{equation}
\label{spect}
S_c(\omega)=-\frac{2\langle c_p\rangle^2}{\omega
  k_BT}\mathrm{Im}\left[\frac{\delta \tilde{F}(\omega)}{\delta \tilde{n}(\omega)}\right].
\end{equation}
We assume that the receptor measures and
averages the concentration over a time $\tau_m$ much longer than any
correlation time. We are thus interested in the low frequency
limit $S_c(\omega\to 0)$ leading to $\overline{\delta
  c_p^2}=S_c(\omega\to 0)/\tau_m$ and hence take
$\Xi(\mathbf{k},\omega)\approx\Xi(\mathbf{k},0)$. 
Finally, cutting the integral in the inverse Fourier
transform $|\mathbf{k}|_{max}= \pi/a$ to correct for a finite target
size (and thereby avoid the UV divergence) we arrive at Eq. (5) in the main text
with
\begin{equation}
\label{exact}
\Lambda(x_a,x_p)=\int_0^1\Big (1+\Big(\frac{a}{\pi x_p
  q}\Big)^{2}\Big[1-\frac{\tan^{-1}(\pi x_a q/a)}{\pi x_a
    q/a}\Big] \Big)^{-1}\! dq. 
\end{equation} 
The integral in Eq. (\ref{exact}) can be evaluated exactly in the
limits when the active excursions are either very short
$x_a/a\ll 1$ or very long $x_a/a\gg 1$ compared to the size of the
target and gives Eqs.~(6) and (7) in the main text.

Now we have to evaluate the equilibrium concentration in the passive
phase in the general case, where signaling is not necessarily transport
controlled. In absence of a directional bias the concentration will be
uniform in both the active and passive phase. Hence we only need to
solve a system of linear equations for $\langle n\rangle$, $\langle
c_p\rangle$ and $\int\langle
c_a\rangle d\Omega$. It can be shown that the steady state concentration of freely diffusing
molecules $\langle c_p\rangle$ is given by 
\begin{equation}
\langle
c_p\rangle=\frac{c_{\mathrm{tot}}-V^{-1}}{2(1+\tau_a/\tau_p)}-\frac{k_{\mathrm{off}}}{2k_{\mathrm{on}}}
+ \left[\left(\frac{c_{\mathrm{tot}}-V^{-1}}{2(1+\tau_a/\tau_p)}-\frac{k_{\mathrm{off}}}{2k_{\mathrm{on}}}
  \right)^2+\frac{c_{\mathrm{tot}}k_{\mathrm{off}}}{k_{\mathrm{on}}(1+\tau_a/\tau_p)}\right]^{1/2},
\label{generalC}
\end{equation} 
where $V$ is the volume of the cell or its domain. For $k_{\mathrm{off}}/
(k_{\mathrm{on}}\langle c_p\rangle)\to0$ we obtain $\langle
c_p\rangle=c_{\mathrm{tot}}/(1+\tau_a/\tau_p)$, where for convenience we have
absorbed the term $1/V$ in the total concentration $c_{\mathrm{tot}}-V^{-1}\to
c_{\mathrm{tot}}$. Expressing the ratio $\tau_a/\tau_p$ in terms of
$x_p$ and $x_a$,
$\tau_a/\tau_p=\frac{D}{|\mathbf{v}|}\frac{x_a}{x_p^2}$ we observe
that $\langle
c_p\rangle$ and hence the relative accuracy along with $x_a$ and $x_p$
in fact depends on the
factor $\frac{D}{|\mathbf{v}|}$ as well.


\begin{thebibliography}{99}
\bibitem{REM} The molecular motor speed $|\mathbf{v}(\Omega)|\sim1\mu\mathrm{m}/
\mathrm{s}$ is widely independent of the particle size \cite{Alberts}.
\bibitem{Arg} A net directional bias can be effectively
treated as a deterministic drift on the ensemble level. The latter
will quantitatively increase the speed but will not interfere with
active focusing.
\end{thebibliography}
\end{document}